\documentclass[lettersize,journal]{IEEEtran}
\usepackage{amsmath,amsfonts}
\usepackage{algorithmic}
\usepackage{algorithm}
\usepackage{array}
\usepackage{textcomp}
\usepackage{stfloats}
\usepackage{url}
\usepackage{verbatim}
\usepackage{graphicx}
\usepackage{cite}
\usepackage{amsmath,amssymb,amsfonts}
\usepackage{algorithmic}
\usepackage{graphicx}
\usepackage{textcomp}
\usepackage{xcolor}
\usepackage{hyperref}
\usepackage{colortbl}
\usepackage{ulem}
\usepackage{listings}
\usepackage{url}
\usepackage{array}
\usepackage{booktabs}

\usepackage{multirow}
\usepackage{makecell}
\usepackage{adjustbox}
\usepackage{ragged2e} 
\usepackage{subfigure}
\usepackage{tcolorbox}
\usepackage{listings}
\usepackage{xcolor}
\usepackage[utf8]{inputenc}

\newcommand{\tool}{\textsc{AutoMT}}

\begin{document}

\title{\tool: A Multi-Agent LLM Framework for Automated Metamorphic Testing of Autonomous Driving Systems}

\author{
\IEEEauthorblockN{
Linfeng Liang\textsuperscript{1}, Chenkai Tan\textsuperscript{2}, Yao Deng\textsuperscript{1}, Yingfeng Cai\textsuperscript{2}, 
T.Y Chen\textsuperscript{3}, Xi Zheng\textsuperscript{1},
}

\IEEEauthorblockA{\textsuperscript{1}\textit{School of Computing, Macquarie University, Australia}} 
\IEEEauthorblockA{\textsuperscript{2}\textit{Automotive Engineering Research Institute, Jiangsu University, China}} 
\IEEEauthorblockA{\textsuperscript{3}\textit{School of Science, Computing and Emerging Technologies, Swinburne University of Technology, Australia}} 
\thanks{\textsuperscript{*}Corresponding author. Email: james.zheng@mq.edu.au}
}


\maketitle
\begin{abstract}
Autonomous Driving Systems (ADS) are safety-critical, where failures can be severe. While Metamorphic Testing (MT) is effective for fault detection in ADS, existing methods rely heavily on manual effort and lack automation. We present \tool, a multi-agent MT framework powered by Large Language Models (LLMs) that automates the extraction of Metamorphic Relations (MRs) from local traffic rules and the generation of valid follow-up test cases. \tool\ leverages LLMs to extract MRs from traffic rules in Gherkin syntax using a predefined ontology. A vision-language agent analyzes scenarios, and a search agent retrieves suitable MRs from a RAG-based database to generate follow-up cases via computer vision. Experiments show that \tool \ achieves up to 5× higher test diversity in follow-up case generation compared to the best baseline (manual expert-defined MRs) in terms of validation rate, and detects up to 20.55\% more behavioral violations. While manual MT relies on a fixed set of predefined rules, \tool \ automatically extracts diverse metamorphic relations that augment real-world datasets and help uncover corner cases often missed during in-field testing and data collection. Its modular architecture—separating MR extraction, filtering, and test generation—supports integration into industrial pipelines and potentially enables simulation-based testing to systematically cover underrepresented or safety-critical scenarios. 
\end{abstract}


\section{Introduction}

  \begin{figure}[ht]
  \centering
  \includegraphics[width=1\linewidth]{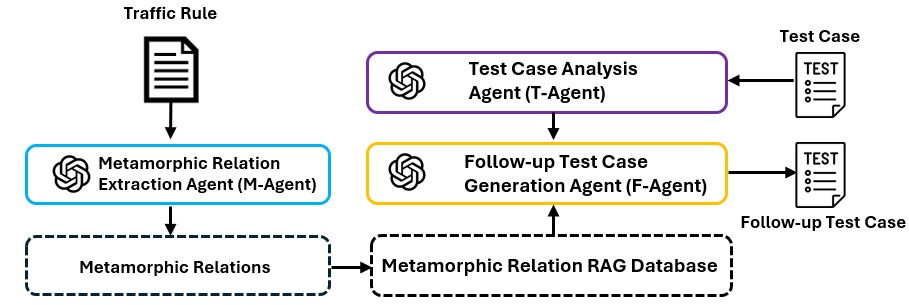}
  \caption{A high-level overview of \tool.}
\label{overview}
\end{figure}
Autonomous Driving Systems (ADSs) are increasingly deployed in real-world settings \cite{caballero2023decision} and are safety-critical, as failures can be catastrophic \cite{zhai2023both, liang2025garl}. Thorough testing is thus essential. Metamorphic Testing (MT) has proven effective for addressing the oracle problem in ADS testing, enabling early bug detection and lowering development costs \cite{tian2018deeptest, zhang2018deeproad, deng2022declarative, yousefizadeh2025using}. Prior work has applied MT to ADSs using real-world data and simulators \cite{yousefizadeh2025using, deng2022declarative}, but significant manual effort hinders full automation. Two key gaps remain: (1) automatically extracting feasible MRs, which prior work relies on human experts for \cite{tian2018deeptest, deng2022declarative}, and (2) identifying suitable original test cases—e.g., a highway test is invalid for the MR \textit{``The vehicle should slow down in a school zone''}. This raises the central question: \textit{How can we automate metamorphic testing for ADSs?}
\par Recent advances in AI, particularly Large Language Models (LLMs), have shown strong performance in language and vision tasks \cite{zhang2025can, shin2024towards, duvvuru2025llm}. We propose \tool, a fully automated MT framework for ADSs powered by multiple LLM-based agents. As shown in Figure~\ref{overview}, \tool\ first uses a predefined ontology based on local traffic rules to extract MRs in Gherkin syntax \cite{dos2018automated} via a MR extraction agent (\textbf{M-Agent}). The extracted MRs are embedded using retrieval-augmented generation (RAG) to populate an MR-RAG database. A vision-language agent (\textbf{T-Agent}) then analyzes test case context (e.g., speed, steering, road type). Finally, a follow-up generation agent (\textbf{F-Agent}) retrieves valid MRs and generates follow-up test cases using computer vision tools. We summarize our key contributions below.

\begin{itemize}

\item \textbf{Automated MR Extraction from Traffic Rules via LLM:} We propose an MR extraction framework that leverages LLMs to automatically extract feasible metamorphic relations from traffic rules, reducing reliance on manual effort and enabling scalable MR discovery.

\item \textbf{RAG-Enhanced Original Test Case Identification:} We introduce a pipeline that combines RAG with LLMs to identify suitable original test cases from real-world datasets or simulators, ensuring the applicability of extracted MRs and enabling valid follow-up test case generation.

\item \textbf{End-to-End Automated Metamorphic Testing:} By integrating automated MR extraction and test case selection, we enable end-to-end automated metamorphic testing, capable of generating follow-up cases for real-world data.

\item \textbf{Extensive Experiments:} We conduct comprehensive experiments, including qualitative analysis of MR extraction accuracy and follow-up case realism, user perception of MR violations in relation to ADS safety, and quantitative comparisons with baseline methods for follow-up test case generation.
\end{itemize}

This paper is organized as follows: Section~\ref{RW} reviews related work. Section~\ref{method} introduces our method, and Section~\ref{experiment} outlines the experimental design. Results are presented in Section~\ref{result}, followed by discussion and future work in Section~\ref{dis}. Section~\ref{con} concludes the paper.

\section{Related Work}
\label{RW}


\subsection{ADS Metamorphic Testing}
The oracle problem poses a major challenge in ADS testing \cite{zhang2023automated}, as real-world traffic complexity makes it infeasible to enumerate all expected outcomes \cite{yousefizadeh2025using}. Metamorphic Testing (MT) \cite{chen2020metamorphic} has been widely adopted to address this issue. Prior work uses a small set of human-defined MRs on real-world datasets, such as DeepTest \cite{tian2018deeptest}, DeepRoad \cite{zhang2018deeproad}, RMT \cite{deng2022declarative}, and MetaSem \cite{yangmetasem}, often applying computer vision tools to detect violations.

Other studies apply MT in simulators \cite{baresi2024efficient, yousefizadeh2025using}. For example, CoCoMEGA \cite{yousefizadeh2025using} uses handcrafted MRs and genetic operators to explore scenario variations, while Baresi et al. \cite{baresi2024efficient} manipulate ego vehicle sensor inputs. However, both real-world and simulation-based approaches rely heavily on manual MR definition and follow-up test case validation \cite{xu2024mr}.

\tool\ automates this process by leveraging LLMs to extract MRs from traffic rules and using a test case analysis agent to verify their applicability to test cases, enabling fully automated MT for ADSs. For simplicity, we focus on real-world data, though the modular design supports simulator-based agents for MR-specific follow-up generation.

\subsection{Automated MR Generation}
Prior studies have explored MR generation. RMT~\cite{deng2022declarative} and MetaSem \cite{yangmetasem} manually parse traffic rules to derive MRs. To reduce manual effort, automated techniques have been proposed. Xu et al.~\cite{xu2025mr} synthesize MRs from test cases, but their method targets structured code testing and lacks generalizability to ADS. Cho et al.~\cite{cho2022automatic} use a genetic algorithm to expand a human-written MR set, while Zhang et al.~\cite{zhang2024scenario} introduce three abstract MR patterns, both still requiring expert input and lacking generality.

LLMs exhibit strong reasoning and generalization abilities \cite{wang2024human}, and have been applied to knowledge extraction \cite{duvvuru2025llm, deng2023target}. Recent work has explored LLM-based MR extraction \cite{zhang2025can, shin2024towards, xu2024mr}. For example, Zhang et al.~\cite{zhang2023automated} generate MRs for ADS modules using LLMs. Others extract MRs from artifacts~\cite{tsigkanos2023large} or requirements~\cite{shin2024towards}. However, due to LLM hallucinations, directly generating MRs often yields many invalid cases.

Our work goes a step further by coordinating multiple specialized agents for fully automated MT. An LLM-based agent extracts MRs from local traffic rules into a RAG database using Gherkin syntax and ontology elements~\cite{dos2018automated, deng2022declarative}. A vision-language agent parses original test cases to match applicable MRs, and a generation agent uses computer vision tools to synthesize valid follow-up test cases based on the matched MRs.

\section{APPROACH}
\label{method}


\subsection{Overview of \tool}


\tool\ comprises three modular components detailed in the following sections. Section~\ref{M-agent} describes the MR extraction agent (M-Agent), which leverages LLMs to construct a retrieval-ready MR database from traffic rules. Section~\ref{Tagent} presents the test case analysis agent (T-Agent), which uses a vision-language model to extract scenario semantics from original test cases. Section~\ref{FAgent} introduces the follow-up generation agent (F-Agent), which retrieves applicable MRs and synthesizes valid follow-up test cases using computer vision techniques.

\subsection{Metamorphic Relation Extraction Agent (M-Agent)}
\label{M-agent}
\begin{figure}[h]
  \centering
  \includegraphics[width=1\linewidth]{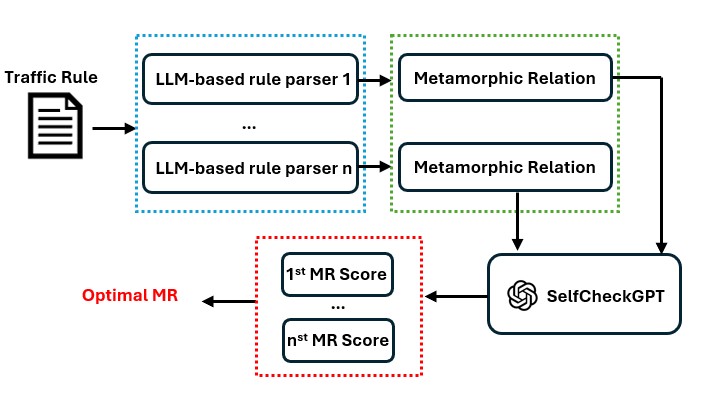}
  \caption{The workflow of M-Agent}
\label{fig_2}
\end{figure}

To enable automated MR extraction, \tool\ introduces the M-Agent. As shown in Figure~\ref{fig_2}, we define a rule parser using Gherkin syntax~\cite{dos2018automated}, a predefined ontology, and LLMs. Given a traffic rule, multiple LLM-based parsers generate candidate MRs, which are validated via SelfCheckGPT~\cite{manakul2023selfcheckgpt} to select the optimal one. The selected MRs are then embedded into a RAG database.




\subsubsection{LLM-based rule parser} 

To systematically extract MRs from traffic rules, we design a structured rule parser based on Gherkin syntax~\cite{dos2018automated}, following the ``Given-When-Then'' pattern commonly used in behavior-driven development (BDD)~\cite{deng2021bmt}. This format maps traffic rule components to three key ontology elements:

\begin{itemize}
    \item \textbf{Given}: \texttt{Road Type} (e.g., intersection, crosswalk, highway)
    \item \textbf{When}: \texttt{Manipulation} (e.g., adding traffic signals, changing weather)
    \item \textbf{Then}: \texttt{Ego-Vehicle Expected Behavior} (e.g., slow down, turn left)
\end{itemize}

This structure ensures that each MR captures a driving context, a scenario transformation, and the expected system response, aligned with traffic regulations.

To define the ontology, we first use an LLM to extract candidate elements from traffic rules using a prompting strategy adapted from Shin~\cite{shin2024towards}, as shown in Table~\ref{tab_Onto}. Then, we refine and categorize them with two ADS experts. Conflicts are resolved by removing duplicates and elements not grounded in the rules. The resulting ontology constrains MR generation, ensuring syntactic validity and semantic fidelity. Table~\ref{ONTOLOGY} presents the summarized ontology categories; the full version is provided in the \href{https://anonymous.4open.science/r/AutoMT-9442/supplementary_material.pdf}{\textbf{Section 1 of the supplementary material}}.

\begin{table}[!t]
\caption{The Prompt for Ontology Elements Extraction}
\begin{tabular}{p{0.95\columnwidth}}
\hline
\rowcolor[HTML]{C0C0C0} 
\textbf{Role Setting}     \\
You are an expert in traffic rules.\\
\hline
\rowcolor[HTML]{C0C0C0} 
\textbf{Prompt}     \\
1. Examples of ontology elements like Table~\ref{ONTOLOGY}.\\
2. I want to derive Road network, Traffic
infrastructure, Object, Environment from a rules document. Can you assist me? Please identify each item that I did not list but is present in the rules document.  \\
User: \textit{One rules document}\\
\hline
\end{tabular}
\label{tab_Onto}
\end{table}



\par For example, consider traffic rule ``Steady Red Light (Stop) Stop before entering the crosswalk or intersection''. The parser would extract the following MR:   

\noindent \textbf{MR example:}
\begin{description}
        \item[Given] the ego-vehicle approaches to an \textit{intersection}
        \item[When] \tool \ \textit{adds} a \textit{ red light on the roadside}
        \item[Then] ego-vehicle should \textit{slow down}
\end{description} 

In this example, the \texttt{Road Type} is \textit{intersection}, defining the driving context. The \texttt{Manipulation} is \textit{adds a red light on the roadside}, representing the scenario change. The \texttt{Ego-Vehicle Expected Behavior} is \textit{slow down}, indicating the ADS should reduce speed in the follow-up scenario (with red light) compared to the source (without red light).


\begin{table}[!t]
\caption{Pre-defined ontology.}
\centering
\resizebox{\linewidth}{!}{
\begin{tabular}{| >{\centering\arraybackslash}m{0.25\linewidth} |
                >{\centering\arraybackslash}m{0.2\linewidth} |
                >{\centering\arraybackslash}m{0.55\linewidth} |}
\hline
\textbf{Category} & \textbf{Level-1 Subcategory} & \textbf{Level-2 Subcategory} \\
\hline
\texttt{Road Type}
    & \makecell{Road network}  & Intersection, Crosswalk, Field path \\
\hline
\multirow{3}{*}{\makecell{\texttt{Manipulation}}}
    & \makecell{Traffic \\infrastructure } &  Sign: STOP Sign etc., Light: Red Light etc., Barriers: Guardrail etc., Line: Crosswalk Markings etc. \\
\cline{2-3}
    & Object  & Vehicle, Pedestrian, Cyclist \\
\cline{2-3}
    & Environment & Rain, Mud, Snowy, Fog, Night \\
\hline
\texttt{Ego-Vehicle
Expected Behavior}  & /   & Slow down, Turn left, Turn right, Keep current \\
\hline
\end{tabular}
}
\label{ONTOLOGY}
\end{table}

\begin{table}[!t]
\caption{The prompt for LLM-based rule parser}
\begin{tabular}{p{0.95\columnwidth}}
\hline
\rowcolor[HTML]{C0C0C0} 
\textbf{Role Setting}     \\
You are an expert in traffic rules and scene analysis. Metamorphic Testing (MT) is a method used in autonomous vehicle testing. Your task is to convert traffic rules into structured "Given-When-Then" metamorphic relations (MRs) for vehicle testing. 

\# Key Concepts \# \\
1. traffic rule: Define how the ego-vehicle should behavioral in the specific driving scenario.\\ 
2. \texttt{Road Type}: Road elements are specified in the traffic rule, such as crosswalk.\\
3. \texttt{Manipulation}: \textit{``adds''}  objects specified in the traffic rule, such as  red light (those items can be added ``on the road'' or ``by the road side'' based on the prior knowledge of LLM), or \textit{``replaces''} environmental conditions, such as a rainy day.\\
4. \texttt{Ego-Vehicle Expected Behavior}: The expected ego-vehicle behavior in the traffic rule, such as slow down, turn right.\\

\# EXAMPLE \# User: Traffic rule: "Steady Red Light (Stop) Stop before entering the crosswalk or intersection"  \\ 
Assistant: Given the ego-vehicle approaches to an intersection\\
When \tool \ adds a red light on the roadside\\
Then ego-vehicle should slow down \\
\hline
\rowcolor[HTML]{C0C0C0} 
\textbf{Prompt}     \\
You are given: \\1. Details of the MRs: ontology elements of  \texttt{Road Type}, \texttt{Manipulation} and \texttt{Ego-Vehicle Expected Behavior}.\\ 
2. To ensure consistency, follow a step-by-step process to extract the MR from traffic rule.\\
Step 1, Determine one appropriate \texttt{Road Type} ontology element based on the rule.
Step 2, Determine one appropriate \texttt{Manipulation} ontology element based on the rule.
Step 3, Determine the verb for \texttt{Manipulation}, use \textit{``adds''}  for objects with optional presence (e.g., pedestrians, vehicles), and \textit{``replaces''} for objects with mandatory presence (e.g., weather, lighting conditions). 
Step 4, Determine one appropriate \texttt{Ego-Vehicle Expected Behavior} ontology element based on the rule.
\\
Finally, compose the MR using the selected elements in the following format: \\Given the ego-vehicle approaches to \texttt{Road Type} \\
When \tool \  \texttt{Manipulation}\\
Then ego-vehicle should \texttt{Ego-Vehicle Expected Behavior}\\
User: \{\textit{One Traffic rule}\} \\

\hline
\end{tabular}
\label{tab_1}
\end{table}

To extract MRs from traffic rules, we adopt a Chain-of-Thought (CoT) prompting strategy~\cite{wei2022chain} to guide LLMs through structured reasoning, as shown in Table~\ref{tab_1}. The process begins by identifying the \texttt{Ego-Vehicle Expected Behavior}, followed by extracting the \texttt{Road Type} and \texttt{Manipulation} elements. These are then composed into a Gherkin-style MR. To improve accuracy, we provide demonstration examples and key concept explanations for in-context learning.



\subsubsection{MR Validation} 
To address LLM hallucination, we adopt SelfCheckGPT~\cite{manakul2023selfcheckgpt}, which assumes that consistent outputs across multiple generations indicate reliable knowledge. As shown in Figure~\ref{fig_2}, we generate multiple candidate MRs per traffic rule using different LLM-based rule parsers. Each output is assessed using a validation prompt with three binary questions (Table~\ref{tab_2}), scoring 0 for ``yes'' and 1 for ``no.'' We compute the average score per MR and select the one with the lowest hallucination score as the final output, ensuring consistency and reliability across LLM generations.



 \begin{table}[!t]
 \caption{The Prompt for MR Validation}
\begin{tabular}{p{0.95\columnwidth}}
\hline
\rowcolor[HTML]{C0C0C0} 
\textbf{Role Setting}     \\
\# CONTEXT \# Based on the list of close-ended yes or no questions, generate a JSON answer. \\
\# Key Concepts \# Same as Table~\ref{tab_1}.\\ 
Questions:\\
1. Are \texttt{Road Type}, \texttt{Manipulation}, and \texttt{Ego-Vehicle Expected Behavior} all mentioned in the traffic rule?\\
2. Is the traffic rule supported by MR?\\
3. Are all parts of the MR consistent with each other? \\
\hline
\rowcolor[HTML]{C0C0C0} 
\textbf{Prompt}     \\
\# EXAMPLE \# User: Traffic rule: "Steady Red Light (Stop) Stop before entering the crosswalk or intersection", \\
MR: "Given the ego-vehicle approaches to an intersection\\
When \tool\ adds a red light on the roadside\\
Then ego-vehicle should slow down"\\
Assistant: ["yes", "yes", "yes"]\\
User: \{\textit{One Traffic rule:}, \textit{MR:}\}  \\  
\hline
\end{tabular}
\label{tab_2}
\end{table}

As shown in Table~\ref{tab_2}, the three validation questions assess different aspects of each MR: (1) coverage of all three ontology elements, (2) alignment with the original traffic rule, and (3) internal and grammatical coherence. Since MRs are synthesized via LLMs using predefined syntax, this step ensures proper integration and correctness.



\subsubsection{Metamorphic Relation RAG Database}



To align test cases with MRs and generate valid follow-ups, we introduce the MR-RAG database. Retrieval-Augmented Generation (RAG) improves LLM performance on domain-specific tasks by grounding outputs in external knowledge~\cite{lewis2020retrieval, fatehkia2024t, zhang2025siren}. We compile extracted MRs into a structured CSV with fields: \texttt{Index}, \texttt{MRs}, \texttt{Road Type}, \texttt{Manipulation}, \texttt{Ego-Vehicle Expected Behavior}, and \texttt{Execution Count}, where the latter tracks MR usage to avoid redundancy. MRs are embedded using OpenAI embeddings~\cite{openai_embeddings} and stored in a FAISS-based vector store~\cite{johnson2019billion}, enabling the F-Agent to retrieve semantically relevant MRs for follow-up test generation.

\subsection{Test Case Analysis Agent (T-Agent)}
\label{Tagent}

Vision-language models (VLMs)~\cite{bai2025qwen2} can act as interactive visual agents. We leverage this capability in the T-Agent to analyze test cases and generate structured representations. These are then matc\-hed with the MR-RAG database by the F-Agent to retrieve relevant MRs. Inspired by DRIVEVLM~\cite{tian2024drivevlm}, the T-Agent is prompted to describe the driving environment—e.g., weather, time, road conditions, and nearby objects—as shown in Table~\ref{tab_3}.



\begin{table}[!t]
\caption{The Prompt for T-Agent}
\begin{tabular}{p{0.95\columnwidth}}
\hline
\rowcolor[HTML]{C0C0C0} 
\textbf{Prompt}     \\
\# Analyze this driving scenario. Describe the time of day, weather conditions, road type (such as intersection, crosswalk, etc.), and any objects around the ego-vehicle. Reply format: time: , weather: , road type: , objects: \\
\hline
\rowcolor[HTML]{C0C0C0} 
\textbf{Images Input}     \\
User: \{\textit{image or a list of images}\}  \\ 
\hline
\end{tabular}
\label{tab_3}
\end{table}

Next, the T-Agent combines its analysis with ground-truth ego-vehicle data (e.g., speed and steering) from the source test case to construct a structured test case representation. This representation is formatted as JSON:


\begin{tcolorbox}[title=Example of test case representation, colback=white!95!gray, colframe=black!80!black, sharp corners, boxrule=0.5pt]
\begin{lstlisting}[basicstyle=\ttfamily\small, breaklines=true]
{
  "Test Case Representation": {
    "Time": "Afternoon",
    "Weather": "Clear",
    "RoadType": "Intersection",
    "Objects": "Cars, buildings, pedestrians, bicycles, trees",
    "EgoVehicle": {
      "Speed": "10.649 km/h",
      "Steering Angle": "-3.689 rad"
    }
  }
}

\end{lstlisting}

\end{tcolorbox}

\subsection{Follow-up Test Case Generation Agent (F-Agent)}
\label{FAgent}
\subsubsection{MR Match}
As shown in Table~\ref{tab_4}, the F-Agent uses the test case representation from the T-Agent to retrieve the most relevant MR from the MR-RAG dataset, prioritizing those with the lowest execution count to reduce repetition. The selected MR's \texttt{Manipulation} element is then used to guide image editing (for real-world data) or scenario modification (in simulation).

\begin{table}[!t]
\caption{The Prompt for F-Agent with MR-RAG Database}
\begin{tabular}{p{0.95\columnwidth}}
\hline
\rowcolor[HTML]{C0C0C0} 
\textbf{Role Setting}     \\
You are an assistant for question-answering tasks. Use the following pieces of retrieved context to answer the question.  \\
\hline
\rowcolor[HTML]{C0C0C0} 
\textbf{Prompt}     \\
\#Given the test case description: \texttt{T-Agent output}, select one MR from the retrieved context where:\\
1. The \texttt{Time}, \texttt{Weather}, \texttt{Road type}, and \texttt{Objects} in the \texttt{T-Agent output} should best match those in the MR.\\
2. Ego-vehicle's speed and steering angle should  match in this MR.\\
3. Among all matched MRs, prefer the one with the lowest \textbf{Execution Count} value.\\
User: \{\textit{test case description}\} 
\\
\hline
\end{tabular}
\label{tab_4}
\end{table}

\subsubsection{Image Editing}
In the \texttt{Manipulation} ontology element, two image operations are defined: ``add'' and ``replace.'' For ``adds,'' F-Agent uses FLUX.1-Fill~\cite{li2024svdqunat}, a state-of-the-art diffusion model designed for localized image edits with high visual fidelity. It accounts for lighting, shadows, and object placement to seamlessly blend modifications into the original image, improving upon prior metamorphic testing techniques~\cite{deng2022declarative,yangmetasem}. FLUX.1-Fill~\cite{flux2024} requires two inputs: an editing prompt and a mask defining the editable region. Masks are generated via semantic segmentation using the OneFormer model~\cite{jain2023oneformer}, fine-tuned on Cityscapes~\cite{cordts2015cityscapes}, which classifies pixels into classes such as person, rider, car, truck, bus, train, motorcycle, and bicycle—key dynamic agents in driving scenarios. For ``replaces,'' F-Agent uses InstructPix2Pix~\cite{brooks2023instructpix2pix}, a text-guided model capable of making global scene edits (e.g., ``make it rainy'') with realistic environmental changes, enabling diverse test scenarios through natural language prompts.

To ensure temporal consistency, we avoid per-frame editing and adopt VISTA~\cite{gao2024vista}, a video generation model trained on driving data that synthesizes realistic sequences using the modified image and vehicle dynamics (e.g., speed). Although test cases generated by \tool\ may not perfectly match real-world fidelity, they preserve critical semantics. Since added elements (e.g., pedestrians, traffic signs) come from real data, unsafe ADS behavior (e.g., failing to slow) remains a valid safety concern.

Our method is modular—image editing tools can be replaced with newer models as they emerge. We use FLUX.1-Fill and InstructPix2Pix due to their strong performance at the time of writing. Additionally, this framework can support test generation in simulation by building libraries that translate matched MRs into scenario scripts.

\section{EXPERIMENT}
\label{experiment}
\subsection{Research Questions}
We evaluate \tool\ through four research questions:  
\textbf{RQ1}: Are the extracted MRs consistent with the underlying traffic rules, as judged by human oracles?  
\textbf{RQ2}: How many MR violations can \tool\ detect across multiple ADSs compared to state-of-the-art baselines?  
\textbf{RQ3}: Are the generated follow-up test cases semantically consistent with their original counterparts?  
\textbf{RQ4}: Do the detected MR violations align with human-perceived safety violations in the tested ADSs?








\subsection{General Experiment Setup}
We evaluate \tool\ on test sets from A2D2~\cite{geyer2020a2d2} and Udacity~\cite{udacity2016ch2}, two large-scale ADS datasets. A2D2 includes 217 cases from Germany (1920$\times$1280), and Udacity contains 167 cases from California (640$\times$480). To standardize and accelerate processing, we downsampled all videos to 10 FPS, which is sufficient for human-perceived fluency~\cite{liang2023cueing}, resulting in 10 frames per test case. To align with regional traffic laws, we manually reviewed official rulebooks~\cite{german_traffic_rules, dmv2025}, identifying 38 distinct rules from Germany and 72 from California (details of those traffic rules can be found \href{https://anonymous.4open.science/r/AutoMT-9442/AutoMT%20Experimental%20Raw%20Results.xlsx}{\textbf{here}}).
For our AI-agents, the MR generation agent (M-Agent) uses three SOTA models: ChatGPT-4o~\cite{hurst2024gpt}, Claude 3.7 Sonnet~\cite{anthropic2025claude37}, and Qwen3-8B~\cite{yang2025qwen3}, with ChatGPT-4o also used for SelfCheckGPT validation. Both the test case analysis agent (T-Agent) and follow-up generation agent (F-Agent) use ChatGPT-4o, a leading vision-language model.

\subsection{Specific Experiment Design}
\subsubsection{RQ1: Qualitative Evaluation of Extracted Metamorphic Relations}
We conducted a qualitative user study with three MT experts to assess whether extracted MRs align with traffic rules. Each expert reviewed all rules and their corresponding MRs and answered: \textit{``Does the metamorphic relation correctly align with the traffic rule?''} Responses were recorded on a 5-point Likert scale. We computed weighted Fleiss' Kappa~\cite{deng2023target} to measure inter-rater agreement by region.

\subsubsection{RQ2: Quantitative Performance of \tool}

As the first fully automated MT pipeline, we compare \tool\ against three adapted baselines based on prior work. All methods take a source test case as input and generate a follow-up test case. Each experiment is run five times, and we report averaged results with statistical analysis. The source code of those baselines can be found in our \href{https://anonymous.4open.science/r/AutoMT-9442/README.md}{\textbf{repo}}.

\begin{itemize}
    \item \textbf{Auto MT Pipeline w/o Traffic Rules~\cite{zhang2023automated}}: Uses ChatGPT-4o to generate an MR directly from a test case without traffic rules or predefined ontology. The model is prompted with MT background and MR structure. Follow-up test cases are generated using the same computer vision tools as \tool.

    \item \textbf{Auto MT Pipeline w/ Traffic Rules~\cite{shin2024towards}}: Uses ChatGPT-4o to generate MRs using traffic rules but without ontology grounding. The model is similarly prompted with MT knowledge and MR format. Follow-up cases are generated using the same tools as \tool.

    \item \textbf{Auto MT with Manually Defined MRs}: To compare with expert-defined MRs, we reviewed prior works~\cite{deng2022declarative, yousefizadeh2025using, yangmetasem} and identified 9 universal, mutually exclusive MRs. For each test case, one MR is randomly selected to generate a follow-up. This design aligns with our approach and other baselines, which generate exactly one follow-up per test case, ensuring a fair and consistent comparison across methods. Random selection also avoids over-representation of particular MRs and reflects realistic scenarios where only one transformation may be applied at a time. Those MRs include adding vehicles, pedestrians, cyclists, and traffic lights or signs, as well as changing the weather to include rainy days, snowy days, and nighttime conditions. 

\end{itemize}

We evaluate follow-up test cases using six ADSs. Four are single-frame models commonly used in MT for ADSs~\cite{deng2022declarative,yangmetasem}:  
\begin{itemize}
\item \textbf{PilotNet}~\cite{bojarski2016end}: NVIDIA's end-to-end driving model,  
\item \textbf{Epoch}~\cite{chrisgundling2017}: a high-performing model from Udacity,  
\item \textbf{ResNet101-ADS}~\cite{deng2022declarative}: based on ResNet101~\cite{he2016deep} with a fully connected output layer,  
\item \textbf{VGG16-ADS}~\cite{deng2022declarative}: similar to above, using VGG16~\cite{simonyan2014very}.
\end{itemize}

Two models process multiple frames:
\begin{itemize}
\item \textbf{CNN-LSTM}~\cite{lai2023end}: a CNN combined with an LSTM and a fully connected layer,  
\item \textbf{CNN3D}~\cite{lai2023end}: a 3D CNN with a fully connected output layer.
\end{itemize}

All ADSs are trained on a combined A2D2 and Udacity dataset with ground truth labels for steering angle (radians) and speed (m/s). We use 80\% for training, 10\% for validation, and 10\% for testing. Images are resized to 320$\times$160 pixels, and training is run to convergence.


For RQ2, we use two metrics: \textbf{Follow-up Test Case Validation Rate} and \textbf{Safety Violation Rate}. The validation rate is the ratio of valid follow-up test cases to the total generated. Validity is determined using three binary evaluation metrics; if any metric is 0, the follow-up test case is considered invalid.

\begin{enumerate}
    \item \textbf{Scenario Alignment:} We check whether the road type remains consistent between the source and follow-up test cases using ChatGPT-4o. A mismatch marks the follow-up as invalid.

    \item \textbf{Logical Alignment:} We assess consistency between the \textbf{Given}, \textbf{When}, and \textbf{Then} components of the MR using SelfCheckGPT. Misaligned logic—e.g., contradictory expected behaviors or context violations—results in an invalid MR. 

    \item \textbf{Manipulation Verification:} We use the Difference Coherence Estimator~\cite{baraldi2025changed} to verify whether the intended Manipulation is visually reflected. Formally:

\end{enumerate}
      \begin{equation}
  \label{eq:coherence}
  C(I_{\text{original}}, I_{\text{follow-up}}, \mathbf{e}) =
  \begin{cases}
  1, & \text{if visual change aligns with } \mathbf{e},\\
  0, & \text{otherwise.}
  \end{cases}
  \end{equation}

where \( \mathbf{e} \) is the manipulation prompt, and \( C(\cdot) \) is computed via a VLM. A return value of 0 marks the follow-up as invalid.

\textbf{Safety Violation Rate} is defined as the proportion of follow-up test cases that trigger MR violations, which we treat as indicators of safety violations in this study. For each test case, ADSs predict frame-level speed and steering; we use the median to mitigate outliers. We then compute prediction variance across all ADSs. If the \texttt{Manipulation} specifies \textit{Slow Down}, the follow-up speed must fall below the variance lower bound. For \textit{Keep Current}, both speed and steering must stay within bounds. For \textit{Turn Left/Right}, steering must fall within the expected directional range. MR Violations are treated as safety-critical behaviors. In RQ4, we further investigate the alignment between MR violations and human-perceived safety violations through a user study.

\subsubsection{RQ3: Consistency of Generated Follow-up Test Cases}

To evaluate whether follow-up test cases are semantically consistent with their corresponding source cases, we conducted a user study on 118 randomly sampled valid test pairs. We recruited 15 licensed drivers (aged 18–60) via Prolific~\cite{Prolific2024}, with diverse demographics (details can be found \href{https://anonymous.4open.science/r/AutoMT-9442/prolific_export_68776254a18ec830e7962fb3.csv}{\textbf{here}}). Each participant was shown a source and follow-up test case and asked: ``How realistic is the follow-up video (e.g., visual quality, consistency with the original video)?'' Responses were recorded on a 5-point Likert scale, with justifications required for ratings below ``Neutral.'' Inter-rater agreement was assessed using weighted Fleiss’ Kappa~\cite{deng2023target}.

\subsubsection{RQ4: Alignment Between MR Violations and Perceived Safety Violations}
Using the same samples and participants as RQ3, we evaluated whether MR violations align with perceived safety. Participants answered: ``Given the transformation, is the ADS’s predicted motion (e.g., speed or steering angle) in the follow-up video reasonable?'' A mismatch between MR-based and user judgment indicates a potential false positive or false negative. Responses used a 5-point Likert scale; ratings above ``Neutral'' suggest the motion was reasonable. Disagreements required justification. Agreement was measured using weighted Fleiss’ Kappa~\cite{deng2023target}.

\section{RESULT}
\label{result}
\subsection{RQ1: Evaluation of MR Extraction}

Table~\ref{tab_6} shows human expert evaluations of MRs extracted by \tool. \textbf{93.3\%} (Germany) and \textbf{94.4\%} (California) of MRs were rated above ``Agree'', demonstrating strong alignment with traffic rules. Fleiss' Kappa scores were \textbf{0.421} (moderate agreement) for Germany and \textbf{0.941} (almost perfect) for California. The lower agreement in Germany may stem from unfamiliar patterns (e.g., St. Andrew’s cross) and translation ambiguity. We further analyze outlier cases with ``Strongly Disagree'' ratings to identify causes of disagreement.



\begin{table}[!t]
\caption{Quantitative result of user study on metamorphic relation generated by \tool.}
\resizebox{\linewidth}{!}{
\begin{tabular}{lccccc}
\toprule
Region & Strongly Agree & Agree & Neutral & Disagree & Strongly Disagree \\
\midrule
Germany & 74.4\% & 18.9\% & 1.1\% & 0\% & 5.6\% \\
California & 91.1\% & 3.3\% & 4.4\% & 1.1\% & 0\% \\
\bottomrule

\end{tabular}
}
\label{tab_6}
\end{table}

\emph{Case Study:}
The traffic rule from Germany that received the first ``Strongly Disagree'' rating states:

\begin{quote}
\textbf{``Maximum Speed Limit Sign: Command or Prohibition. A person driving a vehicle must not exceed the speed limit indicated on the sign.''}
\end{quote}

The generated MR is:

\begin{quote}
\textit{``Given the ego-vehicle approaches any roads, when \tool\ adds a maximum 50km/h speed limit sign on the roadside, then the ego-vehicle should slow down.''}
\end{quote}

However, the user believe that if the vehicle is already traveling below the speed limit, requiring it to slow down further is unreasonable. 

A similar issue arises in the ``Disagree'' case involving a traffic rule from California, which states:

\begin{quote}
\textbf{``A green traffic signal light means GO.''}
\end{quote}

The extracted MR is:

\begin{quote}
\textit{``Given the ego-vehicle approaches an intersection, when \tool\ adds a green traffic signal light on the roadside, then the ego-vehicle should keep current speed.''}
\end{quote}

If the ego-vehicle is stationary, maintaining the same state (i.e., staying still) after the light turns green contradicts the intended rule—it should begin moving instead.

\par These two cases highlight a limitation: an MR may be considered invalid depending on the current status of the ego-vehicle. However, \tool\ effectively addresses this issue by leveraging the MR-RAG database to filter out original test cases that are incompatible with the MR. Specifically, it selects only those scenarios where the vehicle's speed exceeds the posted limit or where the vehicle is not stationary. This demonstrates \tool's capability to handle such contextual nuances effectively. 

\subsection{RQ2: Quantitative Performance of \tool}

\begin{table*}[!t]
\caption{Validation rates and corresponding underlying evaluation metrics of different baselines in Germany and California. Best results are marked in bold.}
\centering
\resizebox{0.95\textwidth}{!}{ %
\begin{tabular}{lcccccc}
\toprule
\multirow{2}{*}{\textbf{Method}} 
& \multicolumn{1}{c}{\multirow{2}{*}{\textbf{Overall Validation Rate}}}
& \multicolumn{3}{c}{\textbf{Underlying Evaluation Metrics}} 
& \multicolumn{1}{c}{\multirow{2}{*}{}} 
& \multicolumn{1}{c}{\multirow{2}{*}{}} \\ 
\\[-0.8em]
\cmidrule(lr){3-5}
& 
& \textbf{Scenario Alignment} 
& \textbf{Manipulation Verification} 
& \textbf{Logical Alignment} 
& 
& \\
\midrule
\multicolumn{7}{c}{\textbf{Germany}} \\
\midrule
\textbf{AUTOMT} & 40.74\% & 98.90\% & 46.36\% & 89.77\% & & \\
\textbf{MT without Traffic Rule} & 36.13\% & 82.58\% & 44.89\% & 97.88\% & & \\
\textbf{MT with Traffic Rule} & 31.52\% & 80.37\% & 39.08\% & 95.30\% & & \\
\textbf{MT with Manual MR} & \textbf{50.41}\% & \textbf{99.17}\% & \textbf{50.60}\% & \textbf{100}\% & & \\
\midrule
\multicolumn{7}{c}{\textbf{California}} \\
\midrule
\textbf{AUTOMT} & \textbf{47.90}\% & 97.72\% & \textbf{50.18}\% & 98.32\% & & \\
\textbf{MT without Traffic Rule} & 32.10\% & 78.44\% & 43.59\% & 91.38\% & & \\
\textbf{MT with Traffic Rule} & 19.64\% & 60.48\% & 32.57\% & 80.72\% & & \\
\textbf{MT with Manual MR} & 38.80\% & \textbf{99.76}\% & 38.92\% & \textbf{100\%} & & \\
\bottomrule
\end{tabular}
}
\label{tab:comparison}
\end{table*}

Table \ref{tab:comparison}
presents the quantitative performance of \tool\ and baseline methods across two regions, showing validation rates and the underlying evaluation metrics. \tool\ outperforms all baselines, including \textbf{MT with Manual MR}, achieving up to 28.26\% improvement in validation rate. Although manual MRs perform slightly better in Germany and in some metrics in California, they are expert-defined and limited in diversity and adaptability. In contrast, \tool\ produces more context-specific and diverse MRs, as highlighted in our diversity analysis.

\begin{figure*}
    \centering
    \includegraphics[width=1\linewidth]{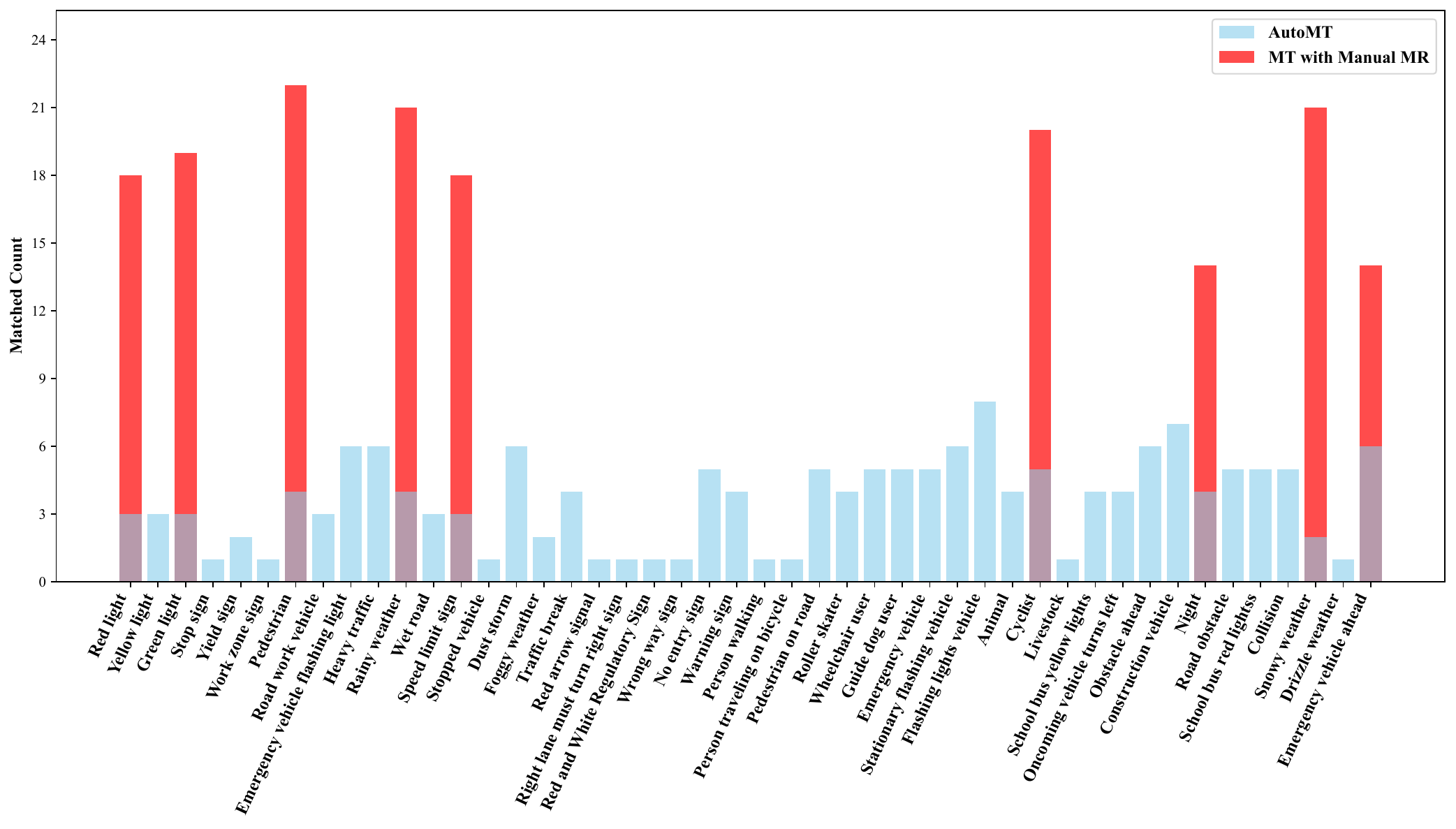}
    \caption{The manipulation operation distribution in the valid follow-up test case generated by \tool \ (blue) and MT with Manual MR (red).}
    \label{mani}
\end{figure*}

Figure~\ref{mani} shows that \tool\ generates 46 distinct manipulations (marked in blue) compared to only 9 in \textbf{MT with Manual MR} (marked in red), demonstrating significantly greater MR diversity. With support from the T-Agent and MR-RAG database, \tool\ more effectively matches MRs to varied test cases, contributing to its superior overall performance.

 The overall validation rate difference between \tool\ and \textbf{MT with Manual MR} is less than 0.6\% across both regions, other than this, \tool\ achieves the highest scores in semantic consistency between follow-up and source test cases. To assess statistical significance, we conducted a t-test on the validation rate. The p-values comparing \tool\ with \textbf{MT without Traffic Rule} and \textbf{MT with Traffic Rule} are as low as 0.002 and 0.0007, respectively, confirming a significant improvement over these baselines.

To understand why baselines struggle with follow-up generation, we analyzed the causes of invalid cases across all methods. A primary issue is failure in the \textbf{Manipulation Verification} metric, largely due to limitations of image editing tools—even the \textbf{MT with Manual MR} baseline is affected. Moreover, the two  baselines show significantly lower pass rates than \tool, indicating that their manipulations are often not properly reflected.

Another major issue is \textbf{Scenario Alignment}, where both baselines perform substantially worse than \tool, which—along with the manual MR baseline—achieves over 98\% alignment. These problems stem from the lack of ontology guidance in baseline MR generation, leading to overly complex MRs that are difficult to execute correctly during editing.




\begin{figure}[ht]
    \centering
    \subfigure[Original test case]{%
        \includegraphics[width=0.45\linewidth]{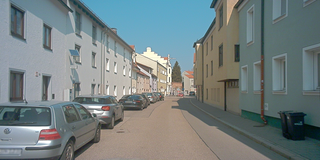}
        \label{fig:case1_ori}
    }\hfill
    \subfigure[Follow-up test case generated by \tool - MR: Given the ego-vehicle approaches to any roads, when method adds a school bus with hazard light on the road, then ego-vehicle should slow down.]{%
        \includegraphics[width=0.45\linewidth]{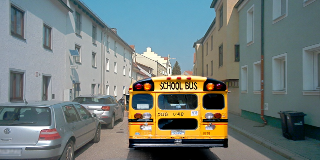}
        \label{fig:case1_auto}
    }\hfill
    \subfigure[Follow-up test case generated by MT with traffic rule - MR: Given the ego-vehicle approaches to a residential street with parked vehicles on both sides, When method adds a pedestrian walking close to the road edge, Then ego-vehicle should slow down.]{%
        \includegraphics[width=0.45\linewidth]{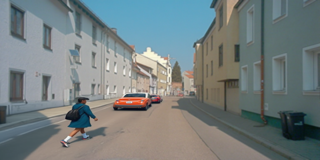}
        \label{fig:case1_other}
    }\hfill
      \subfigure[Follow-up test case generated by MT without traffic rule - MR: Given the ego-vehicle approaches to a narrow urban street with parked cars on both sides, When method adds a pedestrian crossing the street ahead, Then ego-vehicle should slow down]{%
        \includegraphics[width=0.45\linewidth]{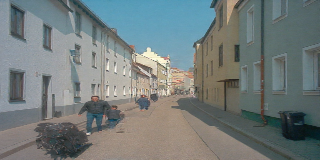}
        \label{fig:case1_other}
}
    
    \caption{Comparison of the original test case, follow-up test case generated by \tool \ and baselines.}
    \label{fig:case_study2}
\end{figure}
Figure~\ref{fig:case_study2} illustrates failure cases from the two baselines, where their MRs do not pass \textbf{Manipulation Verification} and \textbf{Scenario Alignment}. Given the same source test case, \textbf{MT with Traffic Rule} describes the scene as ``a residential street with parked vehicles on both sides'', and \textbf{MT without Traffic Rule} uses ``a narrow urban street with parked cars'', both overly detailed and inaccurate. In contrast, \tool\ produces a concise and accurate description guided by the ontology.

Similarly, the baselines generate overly specific manipulation instructions that image editing tools struggle to execute. \tool\ leverages ontology-based selection to produce clearer, more actionable manipulations, resulting in valid follow-up test cases.


\par For \textbf{Logical Alignment}, both our method and the baselines show high pass rates. However, none reach 100\%, as some MRs are only conditionally valid. Our MR-RAG database helps address this by filtering MRs based on the specific test case context, as also observed in RQ1. Overall, the two LLM-based baselines suffer from hallucinations and overly complex MRs, making it difficult for image editors to generate valid follow-up test cases. This highlights the importance of using predefined ontology elements to guide MR generation and improve validation rates.

\begin{table*}[!t]
\caption{Violation rates of different baselines across six ADSs, each evaluated in Germany and California. Best results are marked in bold.}
\centering
\begin{tabular}{lcccccc}
\toprule
\textbf{Method} 
& \multicolumn{2}{c}{\textbf{PilotNet}} 
& \multicolumn{2}{c}{\textbf{Epoch}} 
& \multicolumn{2}{c}{\textbf{ResNet101-based ADS}} \\
\cmidrule(lr){2-3} \cmidrule(lr){4-5} \cmidrule(lr){6-7}
& Germany & California 
& Germany & California 
& Germany & California \\
\midrule
\textbf{\tool} 
& \textbf{19.08}\% & \textbf{14.37}\%
& \textbf{9.12}\% & \textbf{14.49}\%
& \textbf{25.71}\% & \textbf{21.68}\% \\
\textbf{MT without Traffic Rule} 
& 10.41\% & 7.78\%
& 8.85\% & 9.22\%
& 11.24\% & 11.74\% \\
\textbf{MT with Traffic Rule} 
& 12.17\% & 4.31\%
& 4.33\% & 6.11\%
& 5.16\% & 9.34\% \\
\textbf{MT with manual MR} 
& 13.09\% & 10.54\%
& 4.88\% & 10.66\%
& 11.43\% & 15.69\% \\
\midrule
\textbf{Method} 
& \multicolumn{2}{c}{\textbf{VGG16-based ADS}} 
& \multicolumn{2}{c}{\textbf{CNN-LSTM}} 
& \multicolumn{2}{c}{\textbf{CNN3D}} \\
\cmidrule(lr){2-3} \cmidrule(lr){4-5} \cmidrule(lr){6-7}
& Germany & California 
& Germany & California 
& Germany & California \\
\midrule
\textbf{\tool} 
& \textbf{27.83}\% & \textbf{17.01}\%
& \textbf{16.68}\% & \textbf{12.69}\%
& \textbf{18.16}\% & \textbf{11.62}\% \\
\textbf{MT without Traffic Rule} 
& 9.31\% & 11.86\%
& 9.77\% & 7.07\%
& 13.27\% & 6.35\% \\
\textbf{MT with Traffic Rule} 
& 7.74\% & 6.47\%
& 11.89\% & 3.71\%
& 16.22\% & 3.83\% \\
\textbf{MT with manual MR} 
& 16.22\% & 15.93\%
& 12.53\% & 10.06\% 
& 16.31\% & 8.86\% \\
\bottomrule
\end{tabular}
\label{tab:comparison-violation-ads}
\end{table*}

\par We next examine the violation rate. Table~\ref{tab:comparison-violation-ads} presents the safety violation rates across six ADSs in two regions. \tool\ achieves the highest violation rate in nearly all settings, outperforming baselines by up to 20.55\%. A t-test on the violation rates yields p-values of 0.000193, 0.000096, and 0.000826 when comparing \tool\ with \textbf{MT without Traffic Rule}, \textbf{MT with Traffic Rule}, and \textbf{MT with Manual MR}, respectively—indicating statistically significant differences.

\par
\begin{figure}[ht]
    \centering
    \subfigure[Original test case]{%
        \includegraphics[width=0.48\linewidth,height=0.24\linewidth]{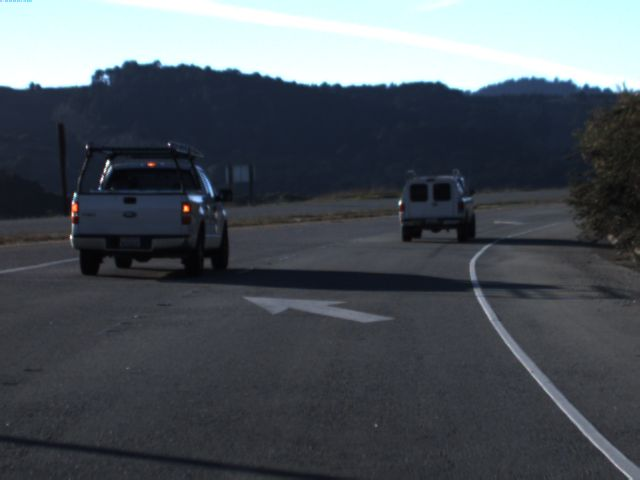}
        \label{fig:case1_ori}
    }
    \hfill
    \subfigure[Follow-up test case generated by MT with manual MR - MR: Given the ego-vehicle approaches to any roads, when method adds a vehicle on the road, then ego-vehicle should slow down.]{%
        \includegraphics[width=0.48\linewidth,height=0.24\linewidth]{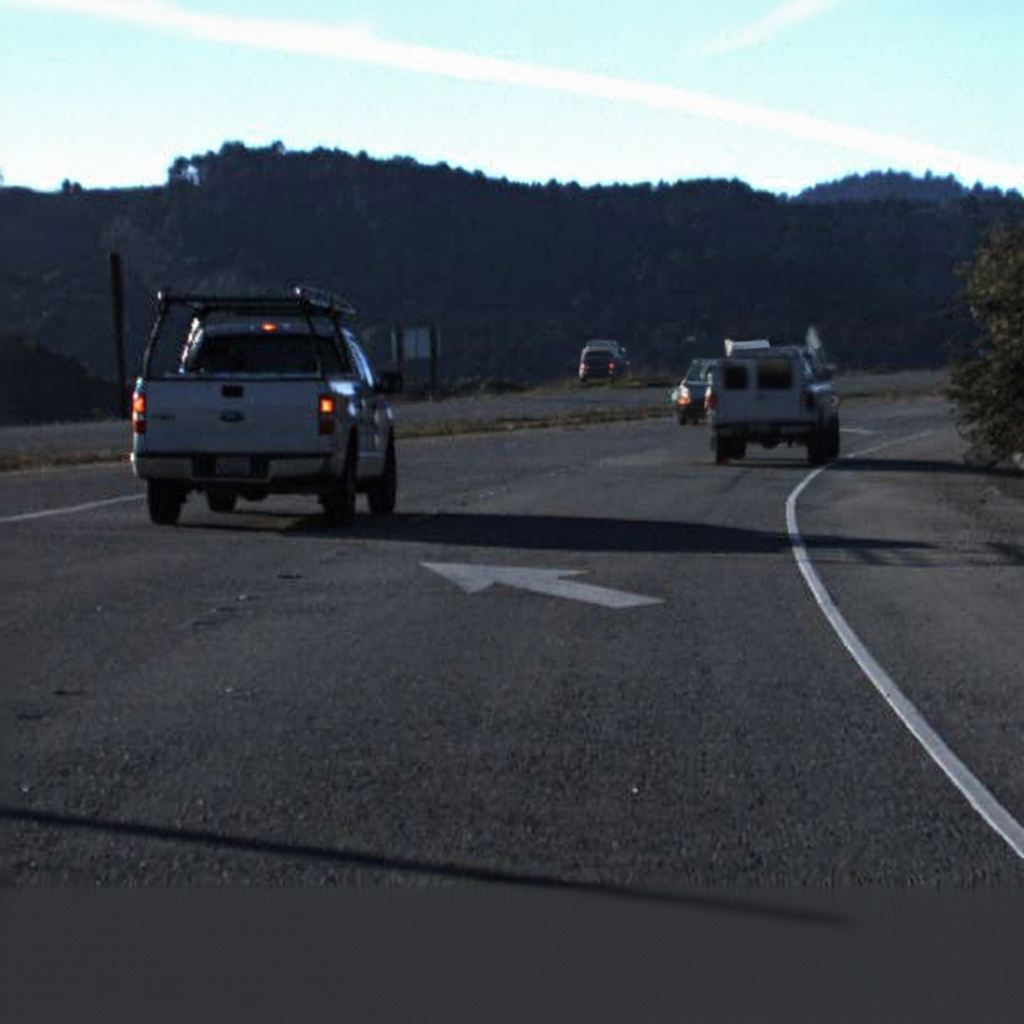}
        \label{fig:case1_manual}
    }

    \vspace{0.5em}

    \subfigure[Follow-up test case generated by \tool\ - MR: Given the ego-vehicle approaches to any roads, when method adds a collision on the road, then ego-vehicle should slow down.]{%
        \includegraphics[width=0.48\linewidth,height=0.24\linewidth]{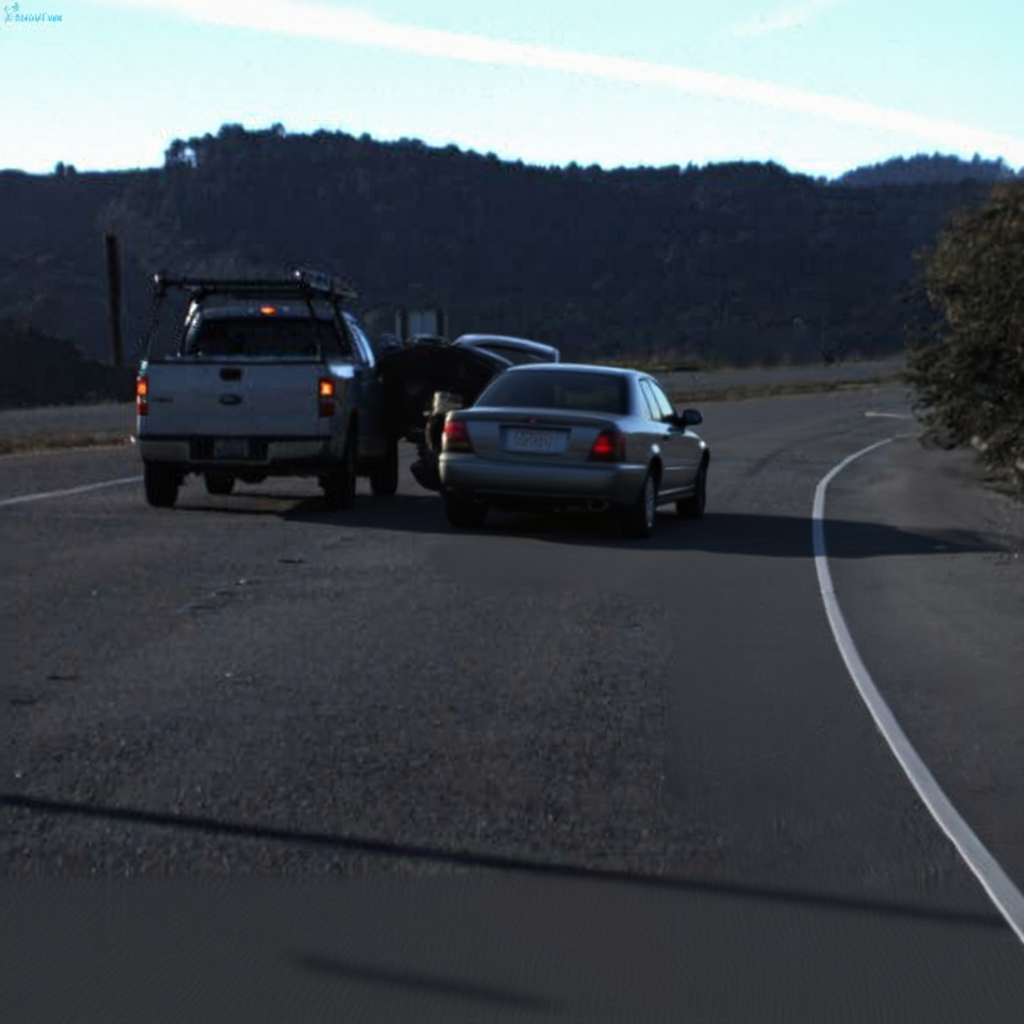}
        \label{fig:case1_tool_collision}
    }
    \hfill
    \subfigure[Follow-up test case generated by \tool\ - MR: Given the ego-vehicle approaches to any roads, when method replaces the weather with a dust storm, then ego-vehicle should slow down.]{%
        \includegraphics[width=0.48\linewidth,height=0.24\linewidth]{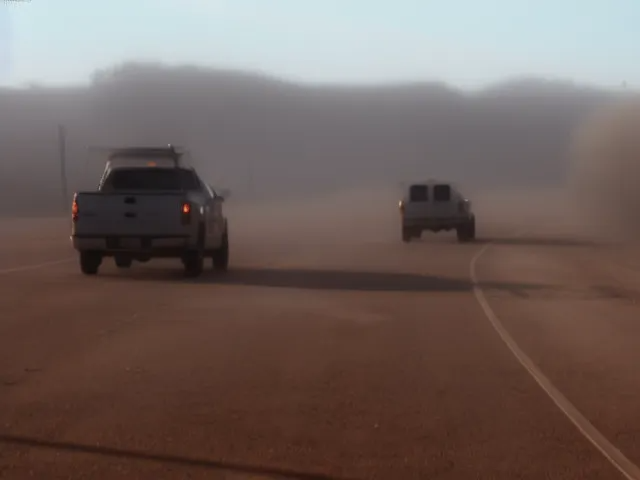}
        \label{fig:case1_tool_weather}
    }

    \caption{Comparison of the original test case and follow-up test cases
    generated by MT with manual MR and \tool.}
    \label{case222}
\end{figure}

 Although \textbf{MT with Manual MR} slightly outperforms \tool\ in validation rate, \tool\ achieves higher violation rates due to a broader range of manipulations (Figure~\ref{mani}). The RAG-based matching aligns MRs more effectively with original test cases, uncovering more diverse unsafe patterns. Some manual MRs may overlap with existing scene elements and thus fail to introduce new violations, as seen in Figure~\ref{fig:case1_manual}. In contrast, \tool\ introduces impactful patterns like collisions or extreme weather (Figures~\ref{fig:case1_tool_collision} and \ref{fig:case1_tool_weather}), which are more likely to trigger violations. This highlights \tool's ability to avoid redundancy and generate diverse, safety-relevant follow-up cases.

\subsection{RQ3: Follow-up Test Case realism Evaluation}


\begin{figure}[ht]
    \centering



    \begin{minipage}[t]{0.48\linewidth}
        \includegraphics[width=\linewidth,height=0.5\linewidth]{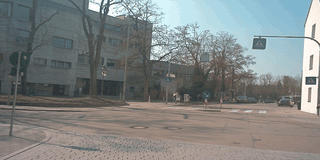}
    \end{minipage}
    \hfill
    \begin{minipage}[t]{0.48\linewidth}
        \includegraphics[width=\linewidth,height=0.5\linewidth]{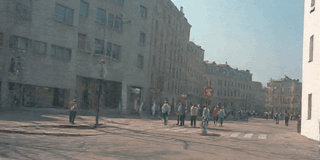}
    \end{minipage}

    \vspace{0.5em}



    \begin{minipage}[t]{0.48\linewidth}
        \includegraphics[width=\linewidth,height=0.5\linewidth]{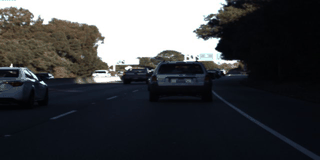}
    \end{minipage}
    \hfill
    \begin{minipage}[t]{0.48\linewidth}
        \includegraphics[width=\linewidth,height=0.5\linewidth]{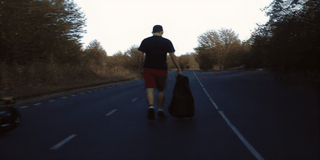}
    \end{minipage}
    
    \vspace{0.5em}
    
    \begin{minipage}[t]{0.48\linewidth}
        \includegraphics[width=\linewidth,height=0.5\linewidth]{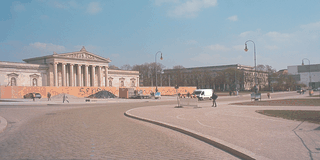}
    \end{minipage}
    \hfill
    \begin{minipage}[t]{0.48\linewidth}
        \includegraphics[width=\linewidth,height=0.5\linewidth]{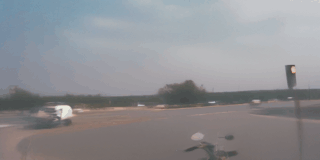}
    \end{minipage}

    \caption{Case study of representative follow-up test cases that are rated as unrealistic by users. For each row, the left image is the original test case, and the right image is the follow-up test case.}
    \label{fig:regional_case_comparison}
\end{figure}

\par To report consistency ratings for 118 follow-up test cases, we summarize descriptive statistics in the main paper and provide two boxplots for user ratings and variance by region (see \href{https://anonymous.4open.science/r/AutoMT-9442/supplementary_material.pdf}{\textbf{Section 8 and Figure 5 \& 6 of the supplementary}}  \href{https://anonymous.4open.science/r/AutoMT-9442/supplementary_material.pdf}{\textbf{material}}). The mean rating is 3.61, median is 4, and 64.5\% of cases are rated above ``Neutral'', indicating most follow-ups are viewed as semantically consistent. Inter-rater agreement via weighted Fleiss’ Kappa is $\kappa = 0.27$, suggesting fair agreement. We also analyze user feedback on low-rated cases, presenting three examples in a case study.


\par Figure~\ref{fig:regional_case_comparison} shows three representative cases rated as unrealistic by users. In the first case,  the required transformation is ``add a group of pedestrians crossing the road''.  \tool\ successfully adds pedestrians, but users noted missing trees, commenting, ``the scenario is not the same as the original.'' The tree removal was caused by FLUX.1-Fill~\cite{flux2024}, which clears space for edits. Since the removed items are non-critical, the test purpose remains valid. Our method is editing-tool agnostic and can readily adopt more advanced tools.



\par The second case involves a transformation that adds a pedestrian to the road. A user deemed it unrealistic, noting that pedestrians rarely appear in such settings. However, while uncommon, this is a valid corner case. \tool\ is designed to surface such rare but plausible scenarios, improving test coverage by uncovering potential safety risks often missed by conventional methods.


\par The third row involves a transformation of adding a yellow roadside light. Users commented that the follow-up appeared blurry and unnaturally smooth, stating it did ``not resemble a real driving scenario.'' Such feedback reflects limitations in current image editing and video generation tools, which is caused by VISTA~\cite{gao2024vista}. This phenomenon is caused by irrational dynamics with respect to historical frames, a common limitation of existing driving world models~\cite{gao2024vista}. These open-source driving world models lack sufficient priors about the future motion tendencies of objects, especially when these objects are generated by diffusion-based models. Despite this, the majority of follow-up test cases were rated realistic, and \tool\ consistently exposed safety violations.

\subsection{RQ4: Safety Violation Validity Evaluation}

\begin{figure}[h]
  \centering
  \includegraphics[width=1\linewidth]{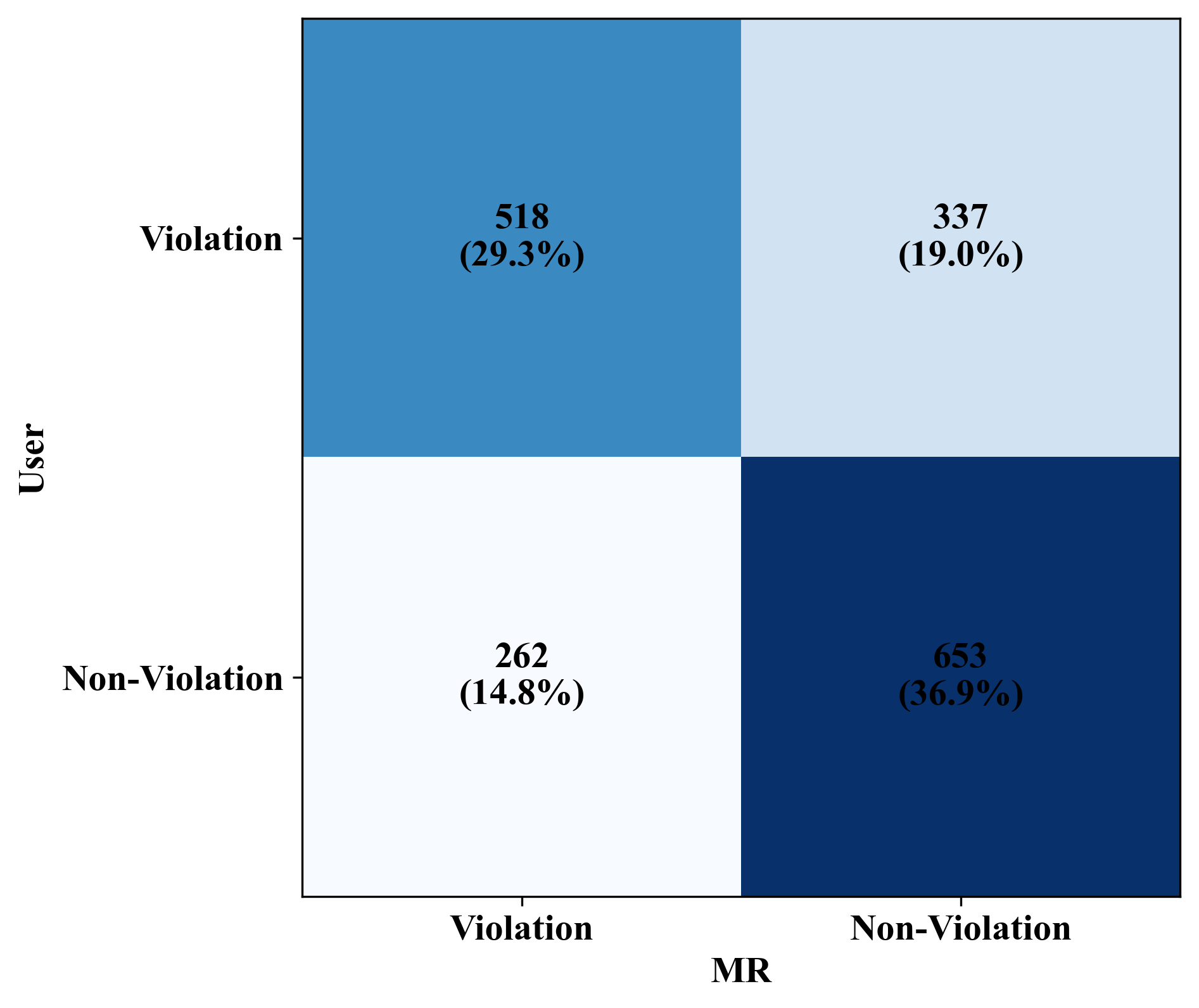}
  \caption{The confusion matrix of safety violation.}
\label{Confusion_matrix}
\end{figure}
\par To assess whether MR violations correspond to real system-level safety issues, we conducted a user study. Figure~\ref{Confusion_matrix} shows the confusion matrix comparing user judgments with MR-based violations. The false positive and false negative rates are 14.8\% and 19\%, while the true positive and true negative rates are 29.3\% and 36.9\%, respectively. The inter-rater agreement, measured by weighted Fleiss’ Kappa, is $\kappa = 0.25$, indicating fair agreement. Overall, users and MRs aligned on 66.2\% of the cases, demonstrating \tool's strong ability to uncover real safety violations.

\begin{figure}[ht]
    \centering

    \begin{minipage}[t]{0.48\linewidth}
        \includegraphics[width=\linewidth,height=0.5\linewidth]{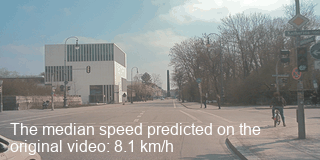}
    \end{minipage}
    \hfill
    \begin{minipage}[t]{0.48\linewidth}
        \includegraphics[width=\linewidth,height=0.5\linewidth]{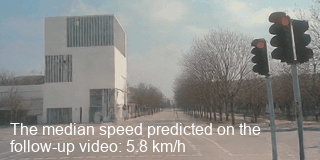}
    \end{minipage}

    \vspace{0.5em}

    \begin{minipage}[t]{0.48\linewidth}
        \includegraphics[width=\linewidth,height=0.5\linewidth]{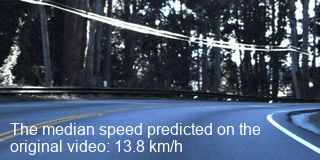}
    \end{minipage}
    \hfill
    \begin{minipage}[t]{0.48\linewidth}
        \includegraphics[width=\linewidth,height=0.5\linewidth]{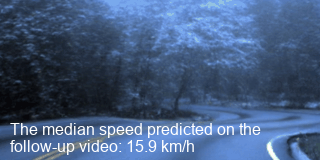}
    \end{minipage}



    \caption{Case study of representative follow-up test cases that exhibit disagreement on whether the case constitutes a violation between the user and MR. For each row, the left image shows the original test case, and the right image shows the follow-up test case.}

    \label{fig:violation_user}
\end{figure}

\par Despite the overall agreement, some disagreements remained. Figure~\ref{fig:violation_user} shows two representative cases. The first row depicts a false negative: the MR was not violated, but users believed a violation occurred. Here, \tool\ added a roadside red light, and although the ego-vehicle slowed down, users expected a complete stop. This reflects a limitation of end-to-end ADSs, which rarely output zero speed due to dataset biases. We plan to address this by extending our pipeline to a simulator and integrating with CoCoMEGA~\cite{yousefizadeh2025using} for more nuanced testing.

\par The second row shows a false positive: the MR is violated, but users perceive no violation. Snowy weather is added, yet the ADS maintains its speed. While MR logic deems this unsafe, users judged the conditions safe enough for steady speed. This discrepancy reflects differing driving habits; however, in principle, reduced speed is expected in snowy conditions to ensure safety. Overall, \tool demonstrates strong capability in revealing real system-level safety violations, even if some edge cases remain subject to human interpretation.

\section{VALIDITY DISCUSSION}
\label{dis}

\subsection{External Validity}
One limitation of our approach is its current focus on offline test case generation using real-world images rather than full-system simulation. While recent work such as CoCoMEGA~\cite{yousefizadeh2025using} explores MT in simulators, our approach reflects the ADS industry's reliance on large-scale in-field data collection, which still suffers from limited scenario diversity~\cite{lou2022testing}. \tool{} complements this by automatically generating diverse test cases through MR-guided augmentation. Our main contribution lies in automated MR extraction from traffic rules, source case analysis for MR applicability, and LLM-based matching via RAG. These modular components can be adapted to simulation-based environments in future work, allowing substitution of image editors with simulation APIs.

\subsection{Internal Validity}
A key internal threat is the reliance on current image editing tools. Even with manual MRs, follow-up test case validation rates remain below 50\% due to rendering limitations. As the CV community advances, we plan to integrate stronger editing backends. Another concern is LLM reliability. Despite using multiple agents across models (GPT-4o~\cite{hurst2024gpt}, Claude~\cite{anthropic2025claude37}, Qwen~\cite{yang2025qwen3}), hallucination remains an issue, as reflected in less-than-perfect logical alignment scores. Ontology constraints and BDD-style syntax help mitigate this issue, further strengthened by our multi-agent collaboration framework. Finally, our violation assessments rely on outputs from VISTA, which may introduce rendering artifacts. We acknowledge this and plan to extend our pipeline to simulation-based testing, using interfaces like CoCoMEGA for higher fidelity.

\subsection{Construct Validity}
We chose not to include other baselines in RQ3 and RQ4 due to three reasons. First RQ2 already compared \tool{} against multiple baselines using three automated metrics—Scenario Alignment, Logical Alignment, and Manipulation Verification—with DCE~\cite{deng2023target} quantifying semantic change. Second, RQ3–RQ4 focus on whether the generated follow-up cases are perceived as realistic and whether MR-predicted violations align with human judgments. These outcomes are influenced more by the image/video rendering tools than by MR quality itself. Including baselines that rely on alternative editing tools would introduce unfair competition and obscure our primary contribution. Third, our technical novelty lies in end-to-end automated MR generation and matching—not in image synthesis—and baseline comparisons in RQ3/RQ4 would misattribute quality differences to the wrong system components.

\section{CONCLUSION}
\label{con}
We propose the first multi-agent framework for extracting MRs from traffic rules with consistency validation, storing them in a RAG-based repository, and using vision-language models to analyze real-world datasets. The pipeline transforms raw videos into structured test case representations and matches them with suitable MRs via reasoning agents guided by diversity and applicability. Extensive experiments—including expert MR validation, testing across multiple ADSs with baselines, assessment of follow-up test case consistency, and alignment of MR violations with human-perceived safety issues—demonstrate the framework's effectiveness. Future work includes extending the modular framework to simulation-based testing and integrating advanced image editing tools to improve diversity coverage for real-world datasets.

\bibliographystyle{IEEEtran}

\bibliography{main}

\begin{thebibliography}{10}
\providecommand{\url}[1]{#1}
\csname url@samestyle\endcsname
\providecommand{\newblock}{\relax}
\providecommand{\bibinfo}[2]{#2}
\providecommand{\BIBentrySTDinterwordspacing}{\spaceskip=0pt\relax}
\providecommand{\BIBentryALTinterwordstretchfactor}{4}
\providecommand{\BIBentryALTinterwordspacing}{\spaceskip=\fontdimen2\font plus
\BIBentryALTinterwordstretchfactor\fontdimen3\font minus \fontdimen4\font\relax}
\providecommand{\BIBforeignlanguage}[2]{{%
\expandafter\ifx\csname l@#1\endcsname\relax
\typeout{** WARNING: IEEEtran.bst: No hyphenation pattern has been}%
\typeout{** loaded for the language `#1'. Using the pattern for}%
\typeout{** the default language instead.}%
\else
\language=\csname l@#1\endcsname
\fi
#2}}
\providecommand{\BIBdecl}{\relax}
\BIBdecl

\bibitem{caballero2023decision}
W.~N. Caballero, D.~Rios~Insua, and D.~Banks, ``Decision support issues in automated driving systems,'' \emph{International Transactions in Operational Research}, vol.~30, no.~3, pp. 1216--1244, 2023.

\bibitem{zhai2023both}
S.~Zhai, S.~Gao, L.~Wang, and P.~Liu, ``When both human and machine drivers make mistakes: Whom to blame?'' \emph{Transportation research part A: policy and practice}, vol. 170, p. 103637, 2023.

\bibitem{liang2025garl}
L.~Liang, Y.~Deng, K.~Morton, V.~Kallinen, A.~James, A.~Seth, E.~Kuantama, S.~Mukhopadhyay, R.~Han, and X.~Zheng, ``Garl: Genetic algorithm-augmented reinforcement learning to detect violations in marker-based autonomous landing systems,'' in \emph{2025 IEEE/ACM 47th International Conference on Software Engineering (ICSE)}.\hskip 1em plus 0.5em minus 0.4em\relax IEEE Computer Society, 2025, pp. 613--613.

\bibitem{tian2018deeptest}
Y.~Tian, K.~Pei, S.~Jana, and B.~Ray, ``Deeptest: Automated testing of deep-neural-network-driven autonomous cars,'' in \emph{Proceedings of the 40th international conference on software engineering}, 2018, pp. 303--314.

\bibitem{zhang2018deeproad}
M.~Zhang, Y.~Zhang, L.~Zhang, C.~Liu, and S.~Khurshid, ``Deeproad: Gan-based metamorphic testing and input validation framework for autonomous driving systems,'' in \emph{Proceedings of the 33rd ACM/IEEE International Conference on Automated Software Engineering}, 2018, pp. 132--142.

\bibitem{deng2022declarative}
Y.~Deng, X.~Zheng, T.~Zhang, H.~Liu, G.~Lou, M.~Kim, and T.~Y. Chen, ``A declarative metamorphic testing framework for autonomous driving,'' \emph{IEEE Transactions on Software Engineering}, 2022.

\bibitem{yousefizadeh2025using}
H.~Yousefizadeh, S.~Gu, L.~C. Briand, and A.~Nasr, ``Using cooperative co-evolutionary search to generate metamorphic test cases for autonomous driving systems,'' \emph{IEEE Transactions on Software Engineering}, 2025.

\bibitem{zhang2025can}
J.~Zhang, C.-a. Sun, H.~Liu, and S.~Dong, ``Can large language models discover metamorphic relations? a large-scale empirical study,'' in \emph{2025 IEEE International Conference on Software Analysis, Evolution and Reengineering (SANER)}.\hskip 1em plus 0.5em minus 0.4em\relax IEEE, 2025, pp. 24--35.

\bibitem{shin2024towards}
S.~Y. Shin, F.~Pastore, D.~Bianculli, and A.~Baicoianu, ``Towards generating executable metamorphic relations using large language models,'' in \emph{International Conference on the Quality of Information and Communications Technology}.\hskip 1em plus 0.5em minus 0.4em\relax Springer, 2024, pp. 126--141.

\bibitem{duvvuru2025llm}
\BIBentryALTinterwordspacing
V.~S.~A. Duvvuru, B.~Zhang, M.~Vierhauser, and A.~Agrawal, ``Llm-agents driven automated simulation testing and analysis of small uncrewed aerial systems,'' in \emph{Proceedings of the IEEE/ACM 47th International Conference on Software Engineering}, ser. ICSE '25.\hskip 1em plus 0.5em minus 0.4em\relax IEEE Press, 2025, p. 385–397. [Online]. Available: \url{https://doi.org/10.1109/ICSE55347.2025.00223}
\BIBentrySTDinterwordspacing

\bibitem{dos2018automated}
E.~C. dos Santos and P.~Vilain, ``Automated acceptance tests as software requirements: An experiment to compare the applicability of fit tables and gherkin language,'' in \emph{Agile Processes in Software Engineering and Extreme Programming: 19th International Conference, XP 2018, Porto, Portugal, May 21--25, 2018, Proceedings 19}.\hskip 1em plus 0.5em minus 0.4em\relax Springer, 2018, pp. 104--119.

\bibitem{zhang2023automated}
Y.~Zhang, D.~Towey, and M.~Pike, ``Automated metamorphic-relation generation with chatgpt: An experience report,'' in \emph{2023 IEEE 47th Annual Computers, Software, and Applications Conference (COMPSAC)}.\hskip 1em plus 0.5em minus 0.4em\relax IEEE, 2023, pp. 1780--1785.

\bibitem{chen2020metamorphic}
T.~Y. Chen, S.~C. Cheung, and S.~M. Yiu, ``Metamorphic testing: a new approach for generating next test cases,'' \emph{arXiv preprint arXiv:2002.12543}, 2020.

\bibitem{yangmetasem}
Z.~Yang, S.~Huang, T.~Bai, Y.~Yao, Y.~Wang, C.~Zheng, and C.~Xia, ``Metasem: metamorphic testing based on semantic information of autonomous driving scenes,'' \emph{Software Testing, Verification and Reliability}, vol.~34, no.~5, p. e1878, 2024.

\bibitem{baresi2024efficient}
\BIBentryALTinterwordspacing
L.~Baresi, D.~Y. Xian~Hu, A.~Stocco, and P.~Tonella, ``{ Efficient Domain Augmentation for Autonomous Driving Testing Using Diffusion Models },'' in \emph{2025 IEEE/ACM 47th International Conference on Software Engineering (ICSE)}.\hskip 1em plus 0.5em minus 0.4em\relax Los Alamitos, CA, USA: IEEE Computer Society, May 2025, pp. 398--410. [Online]. Available: \url{https://doi.ieeecomputersociety.org/10.1109/ICSE55347.2025.00206}
\BIBentrySTDinterwordspacing

\bibitem{xu2024mr}
C.~Xu, S.~Chen, J.~Wu, S.-C. Cheung, V.~Terragni, H.~Zhu, and J.~Cao, ``Mr-adopt: Automatic deduction of input transformation function for metamorphic testing,'' in \emph{Proceedings of the 39th IEEE/ACM International Conference on Automated Software Engineering}, 2024, pp. 557--569.

\bibitem{xu2025mr}
C.~Xu, V.~Terragni, H.~Zhu, J.~Wu, and S.-C. Cheung, ``Mr-scout: Automated synthesis of metamorphic relations from existing test cases,'' \emph{ACM Transactions on Software Engineering and Methodology}, vol.~33, no.~6, pp. 1--28, 2024.

\bibitem{cho2022automatic}
E.~Cho, Y.-J. Shin, S.~Hyun, H.~Kim, and D.-H. Bae, ``Automatic generation of metamorphic relations for a cyber-physical system-of-systems using genetic algorithm,'' in \emph{2022 29th Asia-Pacific Software Engineering Conference (APSEC)}.\hskip 1em plus 0.5em minus 0.4em\relax IEEE, 2022, pp. 209--218.

\bibitem{zhang2024scenario}
Y.~Zhang, D.~Towey, M.~Pike, J.~Cheng~Han, Z.~Quan~Zhou, C.~Yin, Q.~Wang, and C.~Xie, ``Scenario-driven metamorphic testing for autonomous driving simulators,'' \emph{Software Testing, Verification and Reliability}, p. e1892, 2024.

\bibitem{wang2024human}
X.~Wang, H.~Kim, S.~Rahman, K.~Mitra, and Z.~Miao, ``Human-llm collaborative annotation through effective verification of llm labels,'' in \emph{Proceedings of the CHI Conference on Human Factors in Computing Systems}, 2024, pp. 1--21.

\bibitem{deng2023target}
Y.~Deng, Z.~Tu, J.~Yao, M.~Zhang, T.~Zhang, and X.~Zheng, ``Target: Traffic rule-based test generation for autonomous driving via validated llm-guided knowledge extraction,'' \emph{IEEE Transactions on Software Engineering}, vol.~51, no.~7, pp. 1950--1968, 2025.

\bibitem{tsigkanos2023large}
C.~Tsigkanos, P.~Rani, S.~M{\"u}ller, and T.~Kehrer, ``Large language models: The next frontier for variable discovery within metamorphic testing?'' in \emph{2023 IEEE International Conference on Software Analysis, Evolution and Reengineering (SANER)}.\hskip 1em plus 0.5em minus 0.4em\relax IEEE, 2023, pp. 678--682.

\bibitem{manakul2023selfcheckgpt}
P.~Manakul, A.~Liusie, and M.~J. Gales, ``Selfcheckgpt: Zero-resource black-box hallucination detection for generative large language models,'' \emph{arXiv preprint arXiv:2303.08896}, 2023.

\bibitem{deng2021bmt}
Y.~Deng, G.~Lou, X.~Zheng, T.~Zhang, M.~Kim, H.~Liu, C.~Wang, and T.~Y. Chen, ``Bmt: Behavior driven development-based metamorphic testing for autonomous driving models,'' in \emph{2021 IEEE/ACM 6th International Workshop on Metamorphic Testing (MET)}.\hskip 1em plus 0.5em minus 0.4em\relax IEEE, 2021, pp. 32--36.

\bibitem{wei2022chain}
J.~Wei, X.~Wang, D.~Schuurmans, M.~Bosma, F.~Xia, E.~Chi, Q.~V. Le, D.~Zhou \emph{et~al.}, ``Chain-of-thought prompting elicits reasoning in large language models,'' \emph{Advances in neural information processing systems}, vol.~35, pp. 24\,824--24\,837, 2022.

\bibitem{lewis2020retrieval}
P.~Lewis, E.~Perez, A.~Piktus, F.~Petroni, V.~Karpukhin, N.~Goyal, H.~K{\"u}ttler, M.~Lewis, W.-t. Yih, T.~Rockt{\"a}schel \emph{et~al.}, ``Retrieval-augmented generation for knowledge-intensive nlp tasks,'' \emph{Advances in neural information processing systems}, vol.~33, pp. 9459--9474, 2020.

\bibitem{fatehkia2024t}
M.~Fatehkia, J.~K. Lucas, and S.~Chawla, ``T-rag: lessons from the llm trenches,'' \emph{arXiv preprint arXiv:2402.07483}, 2024.

\bibitem{zhang2025siren}
Y.~Zhang, Y.~Li, L.~Cui, D.~Cai, L.~Liu, T.~Fu, X.~Huang, E.~Zhao, Y.~Zhang, Y.~Chen \emph{et~al.}, ``Siren’s song in the ai ocean: A survey on hallucination in large language models,'' \emph{Computational Linguistics}, pp. 1--45, 2025.

\bibitem{openai_embeddings}
\BIBentryALTinterwordspacing
OpenAI, ``Openai embeddings,'' 2024, accessed: 2025-05-27. [Online]. Available: \url{https://platform.openai.com/docs/guides/embeddings}
\BIBentrySTDinterwordspacing

\bibitem{johnson2019billion}
J.~Johnson, M.~Douze, and H.~J{\'e}gou, ``Billion-scale similarity search with gpus,'' \emph{IEEE Transactions on Big Data}, vol.~7, no.~3, pp. 535--547, 2019.

\bibitem{bai2025qwen2}
S.~Bai, K.~Chen, X.~Liu, J.~Wang, W.~Ge, S.~Song, K.~Dang, P.~Wang, S.~Wang, J.~Tang \emph{et~al.}, ``Qwen2. 5-vl technical report,'' \emph{arXiv preprint arXiv:2502.13923}, 2025.

\bibitem{tian2024drivevlm}
X.~Tian, J.~Gu, B.~Li, Y.~Liu, Y.~Wang, Z.~Zhao, K.~Zhan, P.~Jia, X.~Lang, and H.~Zhao, ``Drivevlm: The convergence of autonomous driving and large vision-language models,'' \emph{arXiv preprint arXiv:2402.12289}, 2024.

\bibitem{li2024svdqunat}
M.~Li, Y.~Lin, Z.~Zhang, T.~Cai, X.~Li, J.~Guo, E.~Xie, C.~Meng, J.-Y. Zhu, and S.~Han, ``Svdqunat: Absorbing outliers by low-rank components for 4-bit diffusion models,'' \emph{arXiv preprint arXiv:2411.05007}, 2024.

\bibitem{flux2024}
B.~F. Labs, ``Flux,'' \url{https://github.com/black-forest-labs/flux}, 2024.

\bibitem{jain2023oneformer}
J.~Jain, J.~Li, M.~T. Chiu, A.~Hassani, N.~Orlov, and H.~Shi, ``Oneformer: One transformer to rule universal image segmentation,'' in \emph{Proceedings of CVPR}, 2023, pp. 2989--2998.

\bibitem{cordts2015cityscapes}
M.~Cordts, M.~Omran, S.~Ramos \emph{et~al.}, ``The cityscapes dataset,'' in \emph{CVPR Workshop on the Future of Datasets in Vision}, vol.~2, 2015, p.~1.

\bibitem{brooks2023instructpix2pix}
T.~Brooks, A.~Holynski, and A.~A. Efros, ``Instructpix2pix: Learning to follow image editing instructions,'' in \emph{Proceedings of the IEEE/CVF Conference on Computer Vision and Pattern Recognition}, 2023, pp. 18\,392--18\,402.

\bibitem{gao2024vista}
S.~Gao, J.~Yang, L.~Chen, K.~Chitta, Y.~Qiu, A.~Geiger, J.~Zhang, and H.~Li, ``Vista: A generalizable driving world model with high fidelity and versatile controllability,'' \emph{Advances in Neural Information Processing Systems}, vol.~37, pp. 91\,560--91\,596, 2024.

\bibitem{geyer2020a2d2}
J.~Geyer, Y.~Kassahun, M.~Mahmudi, X.~Ricou, R.~Durgesh, A.~S. Chung, L.~Hauswald, V.~H. Pham, M.~M{\"u}hlegg, S.~Dorn \emph{et~al.}, ``A2d2: Audi autonomous driving dataset,'' \emph{arXiv preprint arXiv:2004.06320}, 2020.

\bibitem{udacity2016ch2}
Udacity, ``Self driving car challenge 2 dataset,'' Available at \url{https://github.com/udacity/self-driving-car/tree/master/datasets/CH2}, 2016.

\bibitem{liang2023cueing}
L.~Liang, Y.~Deng, Y.~Zhang, J.~Lu, C.~Wang, Q.~Sheng, and X.~Zheng, ``Cueing: a lightweight model to capture human attention in driving,'' \emph{arXiv preprint arXiv:2305.15710}, 2023.

\bibitem{german_traffic_rules}
\BIBentryALTinterwordspacing
{Bundesministerium der Justiz}, ``Stra{\ss}enverkehrs-ordnung (stvo),'' 2013. [Online]. Available: \url{https://www.gesetze-im-internet.de/stvo_2013/StVO.pdf}
\BIBentrySTDinterwordspacing

\bibitem{dmv2025}
\BIBentryALTinterwordspacing
{California Department of Motor Vehicles}. (2025) California driver handbook. Accessed: 2025-05-24. [Online]. Available: \url{https://www.dmv.ca.gov/portal/handbook/california-driver-handbook/}
\BIBentrySTDinterwordspacing

\bibitem{hurst2024gpt}
A.~Hurst, A.~Lerer, A.~P. Goucher, A.~Perelman, A.~Ramesh, A.~Clark, A.~Ostrow, A.~Welihinda, A.~Hayes, A.~Radford \emph{et~al.}, ``Gpt-4o system card,'' \emph{arXiv preprint arXiv:2410.21276}, 2024.

\bibitem{anthropic2025claude37}
{Anthropic}, ``Claude 3.7 sonnet,'' \url{https://www.anthropic.com/news/claude-3-7-sonnet}, 2025, accessed: 2025-05-24.

\bibitem{yang2025qwen3}
A.~Yang, A.~Li, B.~Yang, B.~Zhang, B.~Hui, B.~Zheng, B.~Yu, C.~Gao, C.~Huang, C.~Lv \emph{et~al.}, ``Qwen3 technical report,'' \emph{arXiv preprint arXiv:2505.09388}, 2025.

\bibitem{bojarski2016end}
M.~Bojarski, ``End to end learning for self-driving cars,'' \emph{arXiv preprint arXiv:1604.07316}, 2016.

\bibitem{chrisgundling2017}
C.~Gundling, ``cg23,'' \url{https://bit.ly/2VZYHGr}, 2017.

\bibitem{he2016deep}
K.~He, X.~Zhang, S.~Ren, and J.~Sun, ``Deep residual learning for image recognition,'' in \emph{Proceedings of the IEEE conference on computer vision and pattern recognition}, 2016, pp. 770--778.

\bibitem{simonyan2014very}
K.~Simonyan and A.~Zisserman, ``Very deep convolutional networks for large-scale image recognition,'' \emph{arXiv preprint arXiv:1409.1556}, 2014.

\bibitem{lai2023end}
Z.~Lai and T.~Br{\"a}unl, ``End-to-end learning with memory models for complex autonomous driving tasks in indoor environments,'' \emph{Journal of Intelligent \& Robotic Systems}, vol. 107, no.~3, p.~37, 2023.

\bibitem{baraldi2025changed}
L.~Baraldi, D.~Bucciarelli, F.~Betti, M.~Cornia, N.~Sebe, and R.~Cucchiara, ``What changed? detecting and evaluating instruction-guided image edits with multimodal large language models,'' \emph{arXiv preprint arXiv:2505.20405}, 2025.

\bibitem{Prolific2024}
{Prolific}, ``General citation guidelines,'' \url{https://www.prolific.com}, 2024, accessed: September 2025.

\bibitem{lou2022testing}
G.~Lou, Y.~Deng, X.~Zheng, M.~Zhang, and T.~Zhang, ``Testing of autonomous driving systems: where are we and where should we go?'' in \emph{Proceedings of the 30th ACM Joint European Software Engineering Conference and Symposium on the Foundations of Software Engineering}, 2022, pp. 31--43.

\end{thebibliography}

\end{document}